# Pressure-induced semimetal to superconductor transition in a three-dimensional topological material ZrTe$_5$

Yonghui Zhou[1,2,*], Juefei Wu[3,*], Wei Ning[2], Nana Li[4], Yongping Du[3], Xuliang Chen[1,2], Ranran Zhang[2], Zhenhua Chi[1,2], Xuefei Wang[1], Xiangde Zhu[2], Pengchao Lu[3], Cheng Ji[5], Xiangang Wan[3,6], Zhaorong Yang[1,2,6]†, Jian Sun[3,6]†, Wenge Yang[4,5]†, Mingliang Tian[2,6]† & Yuheng Zhang[2,6]

[1] Key Laboratory of Materials Physics, Institute of Solid State Physics, Chinese Academy of Sciences, Hefei 230031, China.
[2] High Magnetic Field Laboratory, Chinese Academy of Sciences, Hefei 230031, China.
[3] National Laboratory of Solid State Microstructures, College of Physics, Nanjing University, Nanjing 210093, China.
[4] Center for High Pressure Science and Technology Advanced Research (HPSTAR), Shanghai 201203, China.
[5] High Pressure Synergetic Consortium (HPSynC), Geophysical Laboratory, Carnegie Institution of Washington, 9700 S Cass Avenue, Argonne, IL 60439, USA.
[6] Collaborative Innovation Center of Advanced Microstructures, Nanjing University, Nanjing 210093, China.

* These authors contributed equally to this work.
†Correspondence and requests for materials should be addressed to Z.Y. (e-mail: zryang@issp.ac.cn) or to J.S. (e-mail: jiansun@nju.edu.cn) or to W.Y. (e-mail: yangwg@hpstar.ac.cn) or to M.T. (e-mail: tianml@hmfl.ac.cn).

**As a new type of topological materials, $ZrTe_5$ shows many exotic properties under extreme conditions. Utilizing resistance and *ac* magnetic susceptibility measurements under high pressure, while the resistance anomaly near 128 K is completely suppressed at 6.2 GPa, a fully superconducting transition emerges surprisingly. The superconducting transition temperature $T_c$ increases with applied pressure, and reaches a maximum of 4.0 K at 14.6 GPa, followed by a slight drop but remaining almost constant value up to 68.5 GPa. At pressures above 21.2 GPa, a second superconducting phase with the maximum $T_c$ of about 6.0 K appears and coexists with the original one to the maximum pressure studied in this work. *In situ* high-pressure synchrotron X-ray diffraction and Raman spectroscopy combined with theoretical calculations indicate the observed two-stage superconducting behavior is correlated to the structural phase transition from ambient *Cmcm* phase to high-pressure *C*2/*m* phase around 6 GPa, and to a mixture of two high-pressure phases of *C*2/*m* and *P*-1 above 20 GPa. The combination of structure, transport measurement and theoretical calculations enable a complete understanding of the emerging exotic properties in three-dimensional topological materials happened under extreme environments.**

Since the first report of topological insulator, an extensive attention in recent years has been focused on newly emergent Dirac materials including topological insulators (TIs)[1,2,3], Dirac semimetals[4,5] and Weyl semimetals[5,6,7] for their unique quantum phenomena. $ZrTe_5$ has been studied for a long time due to its large thermoelectric power[8,9], resistivity anomaly[10,11] and large positive magnetoresistance[12]. Recent theoretical works[13,14] have proposed that single-layer $ZrTe_5$ is a large gap Quantum Spin Hall (QSH) insulator, and the bulk $ZrTe_5$ with stacking of many layers locates in the vicinity of a transition between strong and weak topological insulator. These predictions spark the renewed interest in the investigation of its Dirac and topological characters. Indeed, the magneto-transport experiments[15] have observed the chiral magnetic effect, both angle-resolved photoemission spectroscopy (ARPES)[15] and magneto-infrared spectroscopy[16] study show the electronic structure of $ZrTe_5$ is similar with other three-dimensional (3D) Dirac semimetals like $Na_3Bi$[17,18,19] and $Cd_3As_2$[20,21,22,23,24]. These results suggest that $ZrTe_5$ is a very promising system that hosts topological properties and might help to pave a new way for further experimental studies of topological phase transitions.

As one of the fundamental state parameters, high pressure is an effective clean way to tune lattice as well as electronic states, especially in quantum states[25,26,27]. In this work, by performing resistance and *ac* magnetic susceptibility measurements on $ZrTe_5$ single crystal at various pressures up to 68.5 GPa, semimetal to metal and to superconductor transition were induced at a critical pressure of 6.2 GPa. It was interesting to notice that the occurrence of the metallic phase/or superconductivity at 6.2 GPa is accompanied by the complete suppression of the large resistance peak near 128 K by the pressure. Two superconducting phases were observed. One of them shows a sharp resistance drop with a zero resistance near 3.8 K, which is robust to pressure up to 68.5 GPa. The second one that presents a broad resistance drop but with a finite resistance starting from 6.0 K only exists at a pressure above 21.2 GPa. *In situ* high-pressure synchrotron X-ray diffraction and Raman spectroscopy, combined with theoretical calculations confirm these two pressure-induced superconducting phases correspond to the pressure-induced structural transition of $ZrTe_5$ crystal.

**Results**

**Pressure-induced superconductivity.** The crystals show a thin elongated rectangular shape, where the prismatic $ZrTe_6$ chains run along the crystallographic *a*-axis and linked along the *c*-axis via zigzag chains of Te atoms to form two-dimensional (2D) layers, stacked along the *b*-axis into a crystal[28]. The freshly cleaved crystal was pressurized along *b*-axis and the standard four-probe resistance measurement was performed along the *a*-axis, as illustrated schematically in the inset of Fig. 1a.

Figure 1a and 1c shows the evolution of electrical resistance as a function of temperature for $ZrTe_5$ single crystal at various pressures. Upon cooling at 0.5 GPa from 300 K down to 1.8 K, the overall behavior of resistance displays a typical semiconducting-like feature above 128 K, then the resistance decreases with the decrease of temperature followed by a slight upturn below 20 K. The large resistance anomaly around 128 K is quite similar to those observed under ambient pressure[12,15,29] and was generally correlated to the sign change of charge carriers although the origin still remains elusive[29]. With increasing pressure, the peak temperature increases initially up to 150 K and then shifts back towards lower temperatures accompanied by the broadening of the peak and the decrease of the peak resistance. Since it was reported that the resistance anomaly could be strongly enhanced by an application of a magnetic field at ambient pressure, we investigated the magnetoresistance (MR) at the peak temperature under different pressures. As shown in Fig. 1b, the magneto-transport properties at each pressure show positive MR behavior with the increase of magnetic field but the magnitude of the MR decreases monotonically with the increase of pressure no matter where the peak temperature locates.

Surprisingly, with further increasing pressure, accompanied by the complete suppression of the resistance anomaly peak, a metallic transport behavior with an almost constant normal state resistance within 5% is obtained and a small drop of resistance is observed at ~2.5 K and 6.2 GPa. At 6.7 GPa, the resistance drops to zero at 1.8 K, indicative of the appearance of superconductivity (see Fig. 1c). To make sure that the drop of the resistance was indeed a superconducting transition, we carried out *ac* magnetic susceptibility measurements on $ZrTe_5$ at several pressures up to 9.0 GPa. As seen from Fig. 1d, diamagnetic signal is observed at 7.6 GPa and 9.0 GPa, which is in agreement with the resistance results.

Figure 2a and 2d shows the blow-up of the *R-T* curves at different pressures near the superconducting transition. It was clearly seen that the superconducting transition is quite sharp between 8.3 GPa and 21.2 GPa, indicating the bulk superconductivity.

However, when the pressure is applied up to 30.0 GPa, the onset temperature of the resistance drop occurs from ~6.0 K, accompanied by a gradual decrease of resistance down to 3.9 K, at which the sharp drop of resistance to zero presents, as shown in Fig. 2d. In fact, such a two-step-like transition can be recognized from 25.3 GPa, implying two superconducting phases coexist in the sample. With further increasing pressure, the higher superconducting phase (SC-II) is suppressed gradually but the lower one (SC-I) still survives and remains sharp feature. Clearly, the SC-I exists in a broad pressure regime from 6.2 GPa to the maximum pressure 68.5 GPa achieved in this work, while the SC-II manifests only at pressures above 21.2 GPa.

**Determination of upper critical magnetic field.** Figure 2b and 2e displays temperature dependence of resistance under external magnetic fields aligned along *b*-axis of ZrTe$_5$ at 14.6 GPa and 30.0 GPa, respectively. For both cases, a finite resistive tail at low temperatures was clearly seen and gradually lifted with the increase of magnetic field. A magnetic field of 1.5 T almost smears out the superconducting transition completely. By defining $T_c$ with resistance criterion of $R_{cri} = 90\%R_n$ ($R_n$ is the normal state resistance), we constructed the temperature (*T*) - magnetic field (*H*) phase diagrams, as shown in Fig. 2c and 2f. For *P* = 14.6 GPa, $T_c$ decreases monotonically with increasing magnetic field. The upper critical field $\mu_0 H_{c2}(0)$ is estimated to be about 1.54 T according to the Werthamer-Helfand-Hohenberg (WHH) formula[30]. However, for *P* = 30.0 GPa, being associated with the coexistence of SC-I and SC-II, the *T-H* phase diagram is clearly divided into two parts. The $T_c$ is depressed more pronouncedly at fields below 0.2 T, indicating that the SC-II is sensitive to the magnetic field. At fields above 0.2 T, the superconductivity is dominated by SC-I, where the $R$-$T$ curves are similar to those measured at 14.6 GPa. The $T_c$-$H$ relationship can be described with the WHH equation both at fields below and above 0.2 T, which yields $\mu_0H_{c2}(0)$ value of 0.46 T for SC-II and 1.26 T for SC-I. The large difference of $\mu_0H_{c2}(0)$ indicates that the SC-I and SC-II might have different origins.

**Pressure-superconducting phase diagram.** All the characteristic parameters ($T$ and $P$) are summarized in a $T$-$P$ phase diagram shown in Fig. 3. It can be seen that, with increasing pressure, the peak temperature of the resistance anomaly initially increases up to 150 K and then decreases abruptly. When the peak anomaly disappears at the critical pressure of 6.2 GPa, superconducting phase emerges immediately, indicating a possible quantum critical point (QCP) near 6.2 GPa, below which the sample is semimetal with topological character. Furthermore, if we carefully check the pressure-dependent normal state resistance at 300 K and 10 K (see the inset of Fig. 3), both pressure-induced variations of $R_n$ intuitively follows the $T$-$P$ phase diagram. The slight enhancement of the normal state resistance by the application of pressure up to 2.0 GPa is probably an indication of the pressure-induced competition of the multiband carriers of the sample.

With further increasing pressure above 6.2 GPa, $T_c$ increases monotonically until it reaches a maximum of 4.0 K at 14.6 GPa for the SC-I. The SC-II emerges only above 21.2 GPa with the highest $T_c$ of about 6.0 K. It sounds very clear that the pressure-induced SC-I is a bulk superconductor, while the SC-II without zero resistance is not. There are a number of possibilities that may be responsible for this tiny resistance drop near 6.0 K. One of them is from the impurity or defect phase in the sample. Because the resistance drop reaches about 20%$R_n$, the impurity is almost unlikely in our single crystal sample but the pressure-induced disordering cannot be excluded. The second possibility is due to the occurrence of surface superconductivity associated with the topological surface state, since similar two-step-like transition behavior has also been reported in putative topological superconductor half-Heuslers[31]. Unfortunately, the surface state of $ZrTe_5$ has not been verified yet. The third possible origin is that it is an independent superconducting phase induced by the pressure, but this phase is metastable and coexists with SC-I with different structures.

**The theoretical calculations on the electronic bands and possible crystal structures.** To have a comprehensive understanding of the pressure-induced variations of physical properties of $ZrTe_5$, we also performed the density functional theory calculations for the electronic band structures (see Supplementary Fig. S1). Our results agree well with the previous study, and confirm that this compound is a weak topological insulator at ambient pressure[13]. To get more insights on the high-pressure phases of $ZrTe_5$, we used crystal structure prediction techniques to search the possible candidates under pressure up to 40 GPa. The enthalpy-pressure ($\Delta H$-$P$) curves plotted in Fig. 4 show the best candidates from our structural prediction. The *Cmcm* structure is the most stable one at ambient condition, which is in agreement with the experiments. Upon compression, we find that a layered *C*2/*m* structure (denoted as *C*2/*m*-1) has the lowest enthalpy from 5 GPa to 20 GPa, while an alloy *P*-1 structure is the lowest entropy phases beyond 20 GPa, concurrent with the two-stage superconducting behavior under similar pressure range. The crystal structures of *C*/2*m* and *P*-1 are plotted in Fig. 4b and 4c.

**Structure determination with *in situ* high-pressure synchrotron X-ray diffraction and Raman spectroscopy**. To confirm the structure stability and the predicted new phases, we have conducted *in situ* high-pressure synchrotron X-ray diffraction study on the $ZrTe_5$ sample up to 55 GPa. Angle-dispersive X-ray diffraction experiments were performed on powder $ZrTe_5$ sample at room temperature. A sample chamber in the rhenium gasket pre-indented to 40-μm thickness followed by drilling a 100-μm hole in the center was filled with powder sample, ruby ball and pressure-transmitting medium for *in situ* high-pressure study. The series powder diffraction patterns are shown in Fig. 5a. It is clear to see the structural transitions as pressure increases. Utilizing the GSAS structure refinement[32], the corresponding phases and unit cell volumes fitted by the third-order Birch-Murnaghan equation of state[33] are displayed in Fig. 5b. From these XRD data, it is clear that two structural transitions occurred around 6.0 GPa and above 30 GPa, accompanied with the unit-cell volume drops of ~ 4.4% and ~ 4.8% at these phase transition critical pressures. Rietveld refinements of the high-pressure XRD patterns with known and predicted structures at these typical pressure points are shown in Fig 6. The Bragg peaks in the experimental patterns can be well indexed by the orthorhombic phase (space group *Cmcm*) around ambient pressure and a layered monoclinic phase (*C2/m*) for the pattern at 24.5 GPa. This clearly reveals the semimetal to superconductor transition at around 6 GPa is related to the phase transition from *Cmcm* to *C2/m*. In the high pressure range of 30.9-55.0 GPa, the story seems to be more complicated. Considering the mixture of monoclinic (*C2/m*) and triclinic (*P*-1) phases, the refinement overall matches with the experiment pattern. However, the feature with sharp peaks seems to indicate the occurrence of a cubic symmetry phase. Based on our current observation, we cannot rule out the possibility of separation of body-centered cubic (bcc) Te or a substitutional alloy of $ZrTe_x$ phase in form of any cubic structure in this pressure range.

We further performed Raman scattering measurements under pressure, as shown in Supplementary Fig. S2. $ZrTe_5$ crystal at ambient pressure occupies an orthorhombic *Cmcm* ($D_{2h}^{17}$, No. 63) form, where the Wyckoff positions are 4c for Zr, 4c for Te1, 8f for Te2 and Te3 atoms[28]. Group theory analysis predicts eighteen Raman-active modes in $ZrTe_5$, but only twelve of them were experimentally observed previously, including $B_{1g}$, $B_{2g}$, $B_{3g}$, and $A_g$ modes[34]. In addition, no striking changes in the Raman spectra were observed when the temperature is across the peak temperature in the electrical resistance[35], indicating no structural phase transition[11]. The inset of Supplementary Fig. S2 shows the prominent four $A_g$ modes between 100 and 200 cm$^{-1}$ at ambient pressure, with the most intense peak at 180 cm$^{-1}$, which are consistent with previous reports[34,35,36]. Note that these $A_g$ modes are strongly related to the specific Te2 or Te3 atoms. Under external pressure above 0.6 GPa, the profile of spectra retains similar to that of ambient pressure, whereas the former three modes shift toward higher wavenumber except the mode $A_g^4$, as shown in Supplementary Fig. S2. When the pressure approaches 7.5 GPa where the superconducting transition emerges in *R-T* curve, these modes cannot be detected within the system resolution. These data provide additional evidence of pressure-induced structural phase transition, which matches our theoretical calculations and XRD measurements.

**Discussions**

Our calculations show that the pressure will change the crystal structure dramatically, and at around 6 GPa this compound becomes a metal. As shown in Fig. 4d, 4e and S3, the Fermi surfaces of ZrTe$_5$ at high pressures are very complicated. There are 5 and 7 bands crossing the Fermi level for *C*2/*m* and *P*-1 structure, respectively. A time-reversal-invariant topological superconductor requires odd-parity symmetry and the Fermi surface enclosing an odd number of time-reversal-invariant momenta (TRIM)[37]. The states at Fermi surface are contributed to Zr-4*d* and Te-5*p*. Since these bands are spatially extended, the electronic correlation should be quite small due to the strong screening effect. Consequently, one can expect the superconductivity discovered in this work is mainly mediated by the electron-phonon interaction. Although the intra-pocket phonon-mediated pairing, which may have singular behaviour of the electron-phonon interaction at long wavelengths, can possess odd-parity symmetry[38]. As shown in Fig. 4d, the Fermi surface does not enclose any TRIM. K-point D is not TRIM, although it is enclosed by the red-cyan sheet in Fig. 4d. Thus, this compound is unlikely the topological superconductor at least at low-pressure region (less than 10 GPa). We also perform band structure and Fermi surface calculations for high-pressure region. The Fermi surface becomes quite complicated as shown in Fig. 4e. There are 3 TRIM (A, B, Q) enclosed by two kinds of Fermi surface (denoted by red-cyan and red-blue, respectively.). As a result, the possibility of topological superconductor at higher pressure cannot be exclude, which remains an open question.

From the phonon dispersions of *C*2/*m* and *P*-1 structures at 10 GPa and 30 GPa, respectively, as shown in Fig. S4, both structures are confirmed to be dynamically stable under these pressures. This seems to agree with the experimental results that there are two superconducting phases under pressure. The superconducting properties of *C*2/*m* and *P*-1 structures are not calculated here due to the computational cost. But both of them are metallic phases which can be seen from Fig. 4d, 4e, and S3-S6.

In conclusion, by combining experimental and theoretical investigations, we demonstrated the pressure-induced superconductivity in a Dirac topological semimetal $ZrTe_5$ compound. The appearance of superconductivity at the critical pressure is accompanied by the complete suppression of the high temperature resistance anomaly around 128 K as well as a structural transition from *Cmcm* to *C*2/*m*. At pressures above 21.2 GPa, a second superconducting phase with *P*-1 structure manifests and coexists with the original *C*2/*m*. While our theoretical study rules out the possibility of topological superconductivity at low pressure, at high pressure (above 20 GPa) the system has complicated Fermi surface and a novel superconducting phase, thus deserving further study.

**Methods**

**Single-crystal growth and characterization.** Single crystals of ZrTe$_5$ were grown via vapor transport method in a two-zone furnace with elements Zr (99.99%), Te (99.99%), and iodine (transport agent)[39]. Structural and compositional characterizations of the crystals by X-ray diffraction, scanning (or transmission) electron microscopy, together with electron diffraction and energy dispersive X-ray spectroscopy (EDXS) studies confirm its high quality.

**Experimental details of high-pressure experiments.** High pressures were generated with a screw-pressure-type diamond anvil cell (DAC) made of non-magnetic Cu-Be alloy. The DAC was placed inside a homemade multifunctional measurement system (1.8-300 K, JANIS Research Company Inc.; 0-9 T, Cryomagnetics Inc.) with helium as the heat convection medium. Diamond anvils of 300-μm culets and T301 stainless-steel gasket covered with a mixture of epoxy and fine cubic boron nitride (*c*BN) powder were used for high-pressure transport measurements. Single crystals with typical dimension of 100 × 30 × 10 μm$^3$ were loaded without pressure-transmitting medium. The four-probe method was applied in the *a-c* plane of single crystal. Platinum (Pt) foil with a thickness of 5 μm was used for the electrodes. The high-pressure *ac* magnetic susceptibility was measured using magnetic inductance technique. The 800-μm diamond culets and nonmagnetic Be-Cu gasket were used. Raman scattering measurements were performed at room temperature in a BeCu-type Diacell ST-DAC using a Horiba Jobin Yvon T64000 spectrometer equipped with a liquid nitrogen cooled charge-coupled device. The measurements were conducted in a pseudo backscattering configuration on freshly cleaved single crystal surfaces using 532-nm solid-state laser (torus 532, Laser Quantum) for excitation with a power below 0.1 mW to avoid sample damage and any heating effect. The back-scattered signal was collected in an unpolarized $Y(Z)\bar{Y}$ geometry through 50× objective and 1800 g/mm grating. An integration time of 150 seconds was used. At room temperature, we inserted a flake (typically $100 \times 40 \times 10$ μm$^3$) cleaved parallel to the *a-c* plane and used Daphne 7373 oil as the pressure-transmitting medium. The high-pressure powder X-ray diffraction was conducted at 16-BM-D station, at HPCAT, the Advanced Photon Source, Argonne National Laboratory. A focused monochromatic X-ray beam (~5 microns in FWHM) with wavelength 0.3091 Å was employed for the angle-dispersive diffraction. A Mar345 image plate was used to record two-dimensional diffraction patterns. FIT2D[40] and GSAS[32] programs were used for data integration and structure refinement. Pressure was calibrated by using the ruby fluorescence shift at room temperature for all experiments[41].

**Density functional calculations.** *Ab-initio* random structure searching (AIRSS)[42,43] was applied for crystal structure predictions. Enthalpy calculation was performed using projector augmented wave (PAW) implemented in the Vienna *ab initio* simulation package (VASP)[44]. Phonon dispersion was carried out using finite displacement method with VASP and PHONOPY code[45]. We used the Perdew-Burke-Ernzerhof (PBE) generalized gradient approximation (GGA) exchange-correlation density functional[46]. The Brillouin zone was sampled with a Monkhorst-Pack **k**-mesh with a spacing of 0.03 Å$^{-1}$, and a 2 × 2 × 2 supercell was chosen for phonon calculation. The plane wave cutoff was 288 eV, and structure relaxation was carried out until all the atomic forces on each ion was less than 0.005 eV/Å. Electronic structures were carried out by using full-potential linearized augmented plane-wave (FP-LAPW) method implemented in the WIEN2k package[47]. $63 \times 19 \times 63$ and $63 \times 28 \times 22$ **k**-mesh were used to sample the Brillouin zone for Fermi surface calculation of the structure under 10 GPa and 30 GPa, respectively. Spin-orbit coupling for all elements was taken into account by a second-variation method.

**Acknowledgements**

This research was financially supported by the National Key Projects for Basic Research of China (Grant Nos: 2011CBA00111 and 2015CB921202), the National Natural Science Foundation of China (Grant Nos: U1530402, U1332139, U1332143, U1432251, 11204312, 11374137, 11374302, 51372112 and 11574133), NSF Jiangsu province (No. BK20150012), and the Fundamental Research Funds for the Central Universities. W. Y. acknowledges the financial support from DOE-BES X-ray Scattering Core Program under grant number DE-FG02-99ER45775. Part of the calculations was performed on the supercomputer in the High Performance Computing Center of Nanjing University. HPCAT operations are supported by DOE-NNSA under Award No. DE-NA0001974 and DOE-BES under Award No. DE-FG02-99ER45775, with partial instrumentation funding by NSF. APS is supported by DOE-BES, under Contract No. DE-AC02-06CH11357.

**Figures and captions**

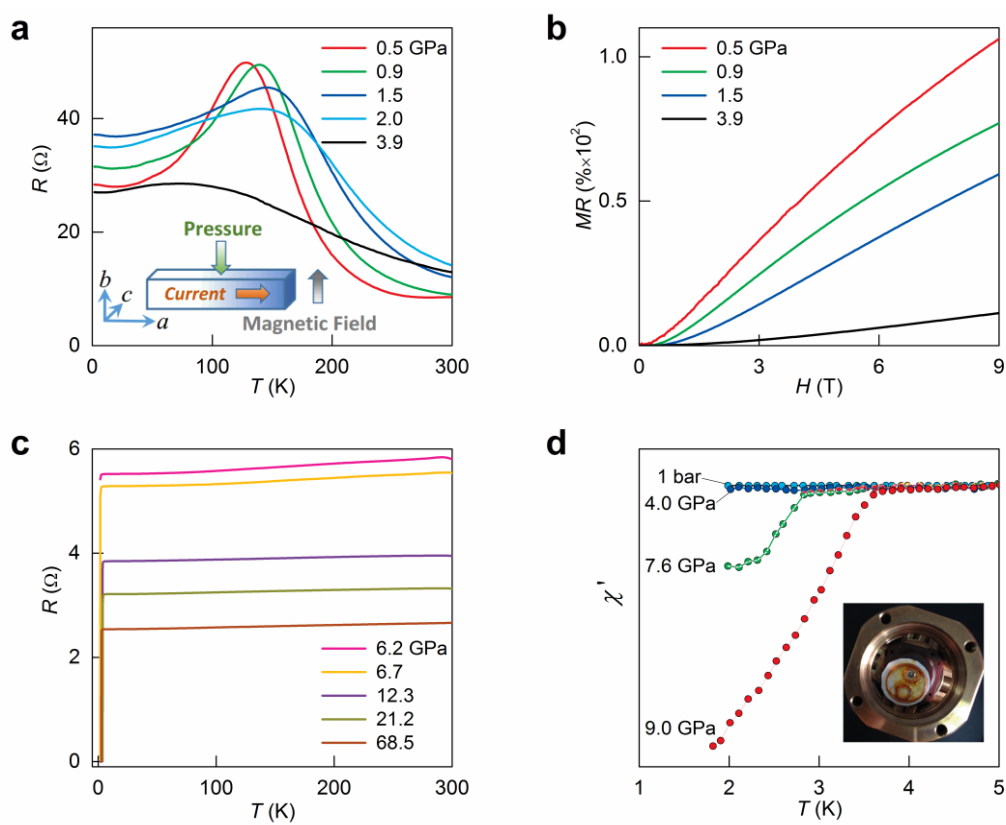

**Figure 1 | Experimental evidence of pressure-induced superconductivity in ZrTe$_5$ single crystal.** (**a**) Temperature-dependent electrical resistance $R(T)$ at various pressures up to 3.9 GPa. The inset schematically shows the arrangements of pressure, magnetic field and current applied. (**b**) Magnetoresistance (MR) measured at the peak temperature of electrical resistance anomaly. The MR is strongly suppressed with increasing pressure. (**c**) The emergence of pressure-induced superconducting transition at higher pressures ranging from 6.2 GPa to 68.5 GPa. (**d**) The real part of the *ac* magnetic susceptibility as a function of temperature at different pressures up to 9.0 GPa. The inset shows the image of experimental setup for the *ac* magnetic susceptibility measurements.

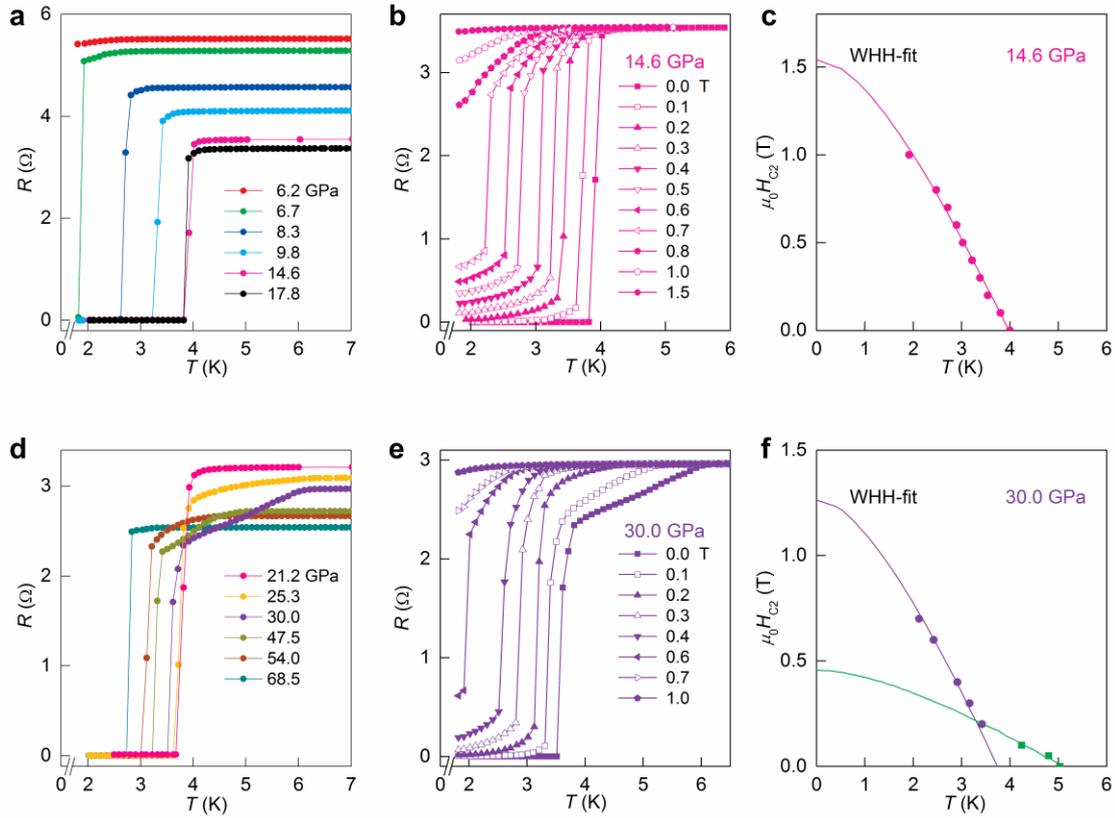

**Figure 2 | Temperature-dependent resistance around superconducting transition temperatures and determination of the upper critical field for the superconducting ZrTe$_5$.** (**a**) The first sharp resistance drop can be seen as the manifestation of superconductivity at 6.7 GPa. With increasing pressure, the $T_c$ increases monotonously towards 17.8 GPa. (**d**) Above 21.2 GPa, a much higher transition emerges at around 6.0 K, suggesting a second superconducting phase. (**b**) **and** (**e**) the temperature dependence of resistance under different magnetic fields parallel to the *b*-axis up to 1.5 T at 14.6 GPa and up to 1.0 T at 30.0 GPa, respectively. (**c**) **and** (**f**) the temperature dependence of the upper critical field $\mu_0 H_{c2}$ at 14.6 GPa and 30.0 GPa, respectively. Here, $T_c$ at different magnetic fields are determined as 90% drop of the normal state resistance. The solid lines represent the fitting lines based on the Werthamer-Helfand-Hohenberg (WHH) formula.

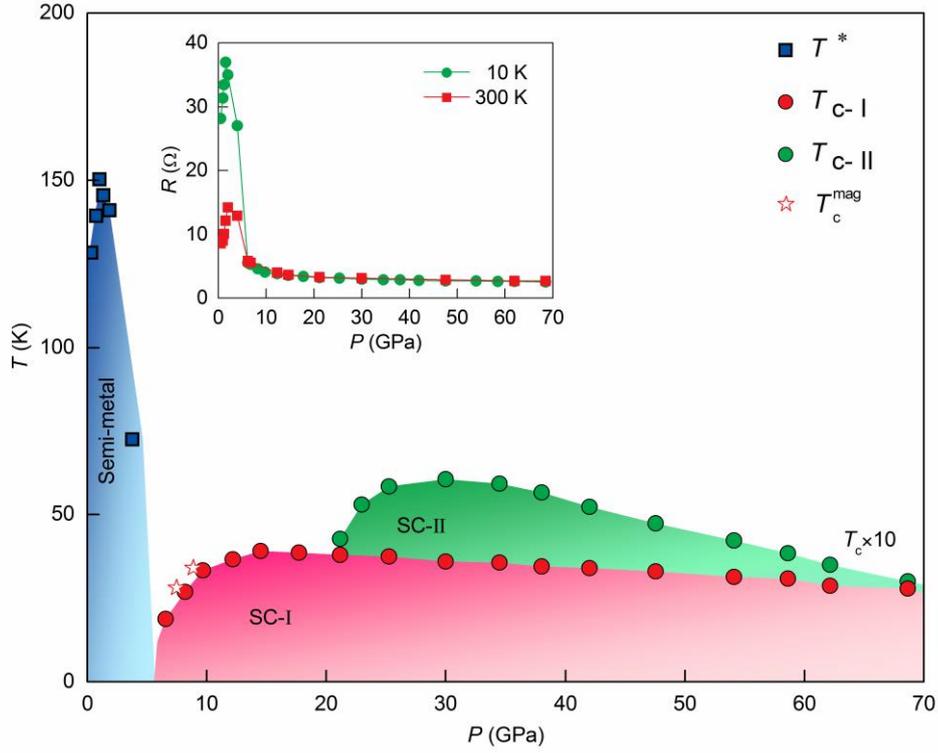

**Figure 3 | Temperature-pressure phase diagram of ZrTe₅ single crystal.** The blue squares denote $T^*$, the peak temperature of electrical resistance anomaly. The solid circles represent $T_c$ extracted from electrical resistance measurements. The pentagrams represent the onset temperature of Meissner effect in the *ac* magnetic susceptibility measurements. Colored areas are a guide to the eyes indicating the distinct phases. The red region corresponds to SC-I phase where $T_c$ is defined as 50% drop of the normal state resistance, while the green region corresponds to SC-II phase where the $T_c$ is determined as the onset temperature of electrical resistance drop. For clarity, the value of $T_c$ is amplified by a factor of 10. The inset shows the specific resistance as a function of applied pressure at 10 K and 300 K, respectively.

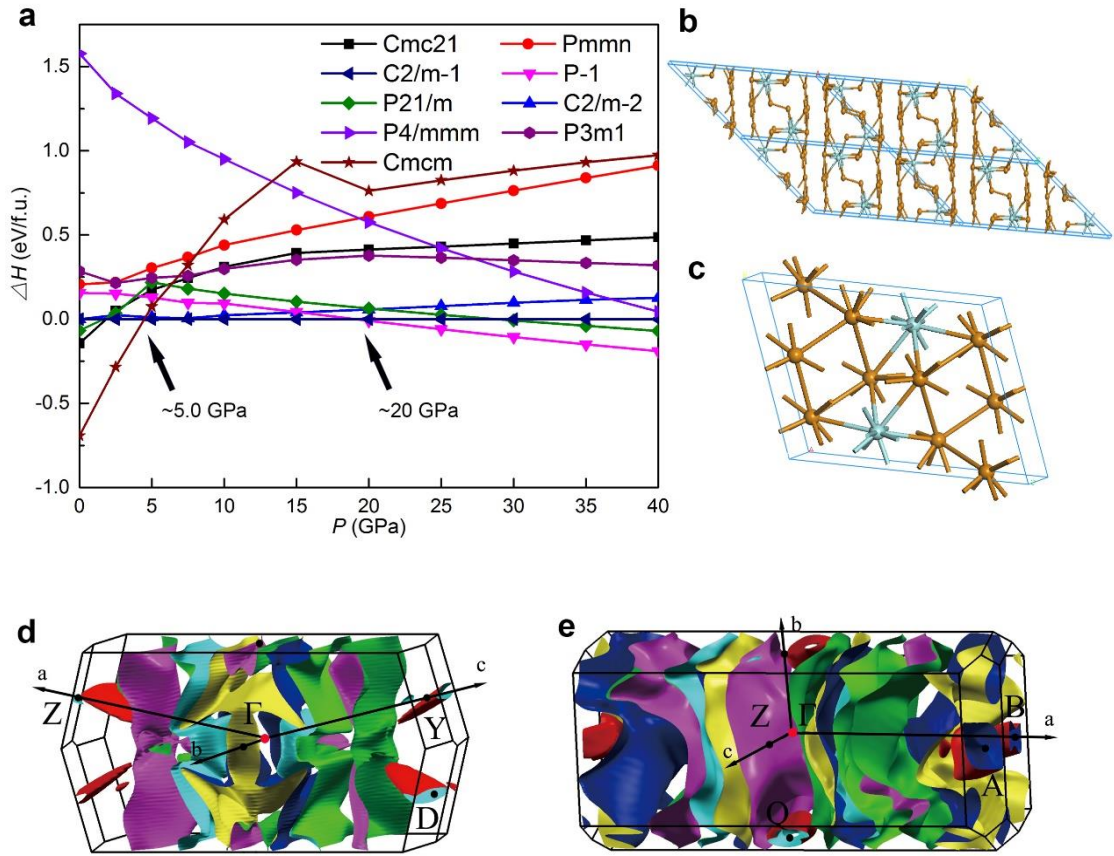

**Figure 4 | Enthalpy calculation of possible stable phases and their atomic and electronic structures of ZrTe$_5$.** Enthalpy vs. pressure for ZrTe$_5$ phases up to 40 GPa (**a**). Crystal structure of *C*2/*m* (**b**) and *P*-1 phase (**c**). The balls in cyan color and golden color represent Zr and Te atoms, respectively. (**d**) Fermi surfaces of *C*2/*m* structure at 10 GPa. (**e**) Fermi surfaces of *P*-1 structure at 30 GPa.

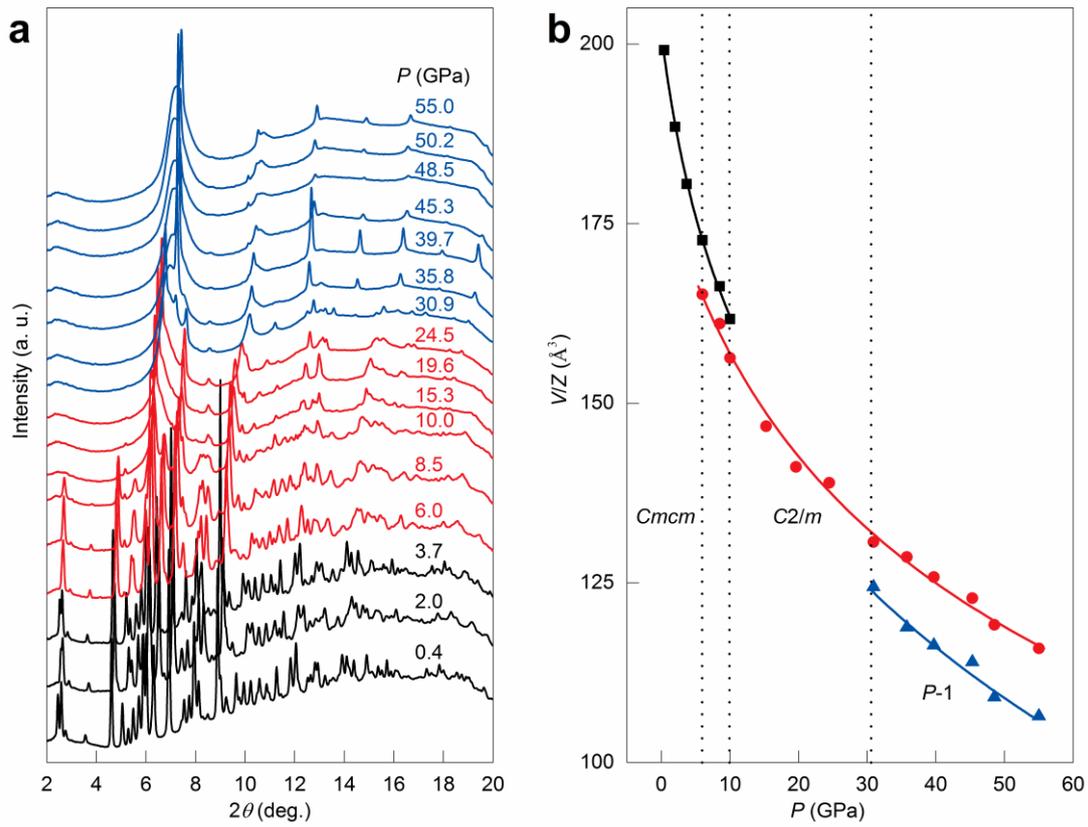

**Figure 5 | High-pressure synchrotron X-ray diffraction patterns of ZrTe$_5$.** (**a**) Representative diffraction patterns at high pressure from 0.4 GPa to 55.0 GPa and room temperature. (**b**) Unit-cell volume per formula unit (*V*/*Z*) as a function of pressure. The solid square, circle, and upper trigonal denote the orthorhombic (*Cmcm*, *Z* = 4), monoclinic (*C*2/*m*, *Z* = 4), and triclinic (*P*-1, *Z* = 2) phase, respectively. The solid lines are the fitting results based on third-order Birch-Murnaghan equation of state. The vertical dot lines are guides for the eyes.

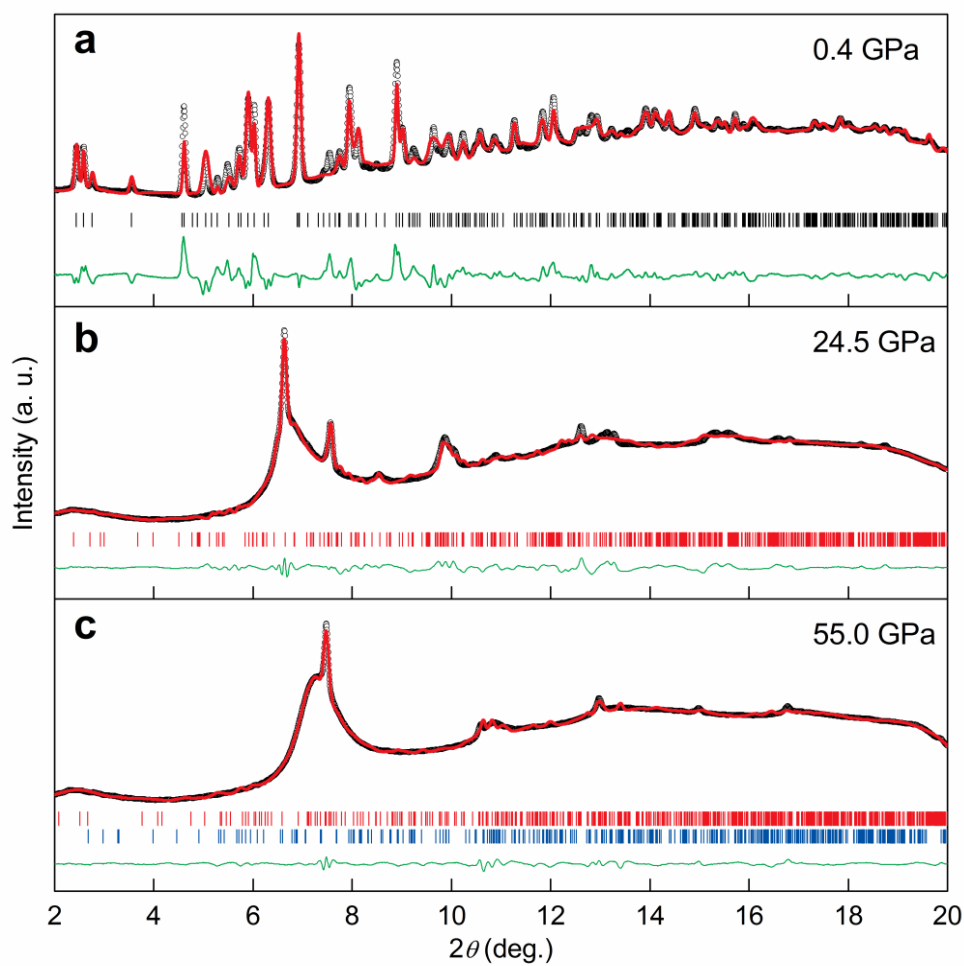

**Figure 6 | Rietveld refinements of the experimental X-ray powder patterns of ZrTe$_5$ with known and predicted low enthalpy structures.** The refinement of the diffraction patterns at (**a**) 0.4 GPa, (**b**) 24.5 GPa, and (**c**) 55.0 GPa with an orthorhombic phase (*Cmcm*), monoclinic phase (*C*2/*m*), and a combination of monoclinic (*C*2/*m*) and triclinic (*P*-1) phases, respectively. The solid lines and open circles represent the Rietveld fits for the lattice and observed data, respectively. The solid lines at the bottom denote the residual intensities. The vertical bars indicate the peak positions.